\documentclass[useAMS,usenatbib,usegraphicx]{mn2e}

\usepackage{times}

\newcommand{\rsun}{\,\mbox{$R_{\odot}$}}
\newcommand{\msun}{\,\mbox{$M_{\odot}$}}

\newcommand{\kms}{\hbox{km s$^{-1}$}}

\newcommand{\degs}{$\degr$}
\newcommand{\chisq}{$\chi^{2}$}

\begin{document}

\title[Differential rotation of HD 155555]{Differential rotation on both components of the pre main-sequence binary system HD 155555}

\makeatother
\author[N.J.~Dunstone et al.]
{N.J.~Dunstone,$^1$\thanks{E-mail: njd2@st-andrews.ac.uk} G.A.J. Hussain,$^{1,2}$ A. Collier Cameron,$^1$ S.C. Marsden,$^3$ M. Jardine,$^1$ 
\newauthor
 J.R. Barnes,$^4$ J.C. Ramirez Velez,$^5$ J.-F.~Donati$^6$ \\
$^1$ School of Physics and Astronomy, University of St Andrews, Fife KY16 9SS, UK\\
$^2$ ESO, Karl-Schwarzschild-Strasse 2, D-85748 Garching, Germany \\
$^3$ Anglo-Australian Observatory, PO Box 296, Epping, NSW 1710, Australia \\
$^4$ Centre for Astrophysics Research, University of Hertfordshire, Hertfordshire AL10 9AB\\
$^5$ LESIA, Observatoire de Meudon, 92195 Meudon, France \\
$^6$ LATT, CNRS-UMR 5572, Obs. Midi-Pyrénées, 14 Av. E. Belin, F-31400 Toulouse, France\\}

\date{2008, 200?}

\maketitle

\begin{abstract}

We present the first measurements of surface differential rotation on a pre-main sequence binary system. Using intensity (Stokes I) and circularly polarised (Stokes V) timeseries spectra, taken over eleven nights at the Anglo-Australian Telescope (AAT), we incorporate a solar-like differential rotation law into the surface imaging process. We find that both components of the young, 18 Myr, HD 155555 (V824 Ara, G5IV + K0IV) binary system show significant differential rotation. The equator-pole laptimes as determined from the intensity spectra are 80 days for the primary star and 163 days for the secondary. Similarly for the magnetic spectra we obtain equator-pole laptimes of 44 and 71 days respectively, showing that the shearing timescale of magnetic regions is approximately half that found for stellar spots. Both components are therefore found to have rates of differential rotation similar to those of the same spectral type main sequence single stars. The results for HD 155555 are therefore in contrast to those found in other, more evolved, binary systems where negligible or weak differential rotation has been discovered. We discuss two possible explanations for this; firstly that at the age of HD 155555 binary tidal forces have not yet had time to suppress differential rotation, secondly that the weak differential rotation previously observed on evolved binaries is a consequence of their large convection zone depths. We suggest that the latter is the more likely solution and show that both temperature and convection zone depth (from evolutionary models) are good predictors of differential rotation strength. Finally, we also examine the possible consequences of the measured differential rotation on the interaction of binary star coronae.

\end{abstract}


\section{Introduction}
\protect\label{sect:intro}

Differential rotation is a key factor in the generation of stellar magnetic fields. In the Sun it is differential rotation that transforms a large-scale polodial field into a stronger toroidal component. In the solar case this is thought to occur in the thin interface layer between the radiative core and the convective envelope called the tachocline. Recent Zeeman Doppler imaging (ZDI) studies of cool stars have found that this may not be the case for all stars. Regions of strong surface azimuthal magnetic field have been observed. The surface field of the Sun consists mainly of radial field and classical dynamo models suggest that any large-scale azimuthal field should be restricted to the tachocline. This has led to the suggestion that the dynamo in young rapidly rotating stars may not be restricted to the base of the convection zone as in the Sun but be distributed throughout (\citealt{donati97recon}). It is therefore important to measure stellar differential rotation to see if this provides any clues to explain these observations.

Techniques for measuring the strength of differential rotation by tracking the cool spots or magnetic regions have been developed from the results of Doppler imaging and Zeeman Doppler imaging (ZDI) studies. So far differential rotation has now been measured for more than 20 stars. These include stars of a range of different surface temperatures and so convection zone depths. \cite{barnes05} collates many of these results, finding that differential rotation rate is strongly related to surface temperature but only weakly correlated with rotation rate. Near fully convective M-dwarfs (e.g. HK Aqr) exhibit very weak differential rotation, while early G-dwarfs have strong differential rotation.

The stars considered in the \cite{barnes05} study were all single young rapid rotators. Measurements of differential rotation have also been made for binary systems, again using Doppler imaging. The limited published sample currently contains the evolved primary star of the RS Cvn systems HR1099 (\citealt{petit04hr1099}) and IM Peg (\citealt{marsden07}) and the main-sequence secondary star of the pre-cataclysmic variable star V471 Tau (\citealt{hussain06}). All three stars are found to have weak differential rotation and in the case of V471 Tau it was consistent with solid body rotation. So the question remains whether it is the evolved nature of the stars or their membership of a binary system that is responsible for the observed decrease in differential rotation strength. In an attempt to disentangle these two effects we observed HD 155555 (V824 Ara), a pre-main sequence binary system consisting of two nearly equal mass components (G5IV + K0IV). 

In total eleven nights of observations were obtained of HD 155555 (as detailed in \S \ref{sect:obs}). A five night subset of the data was used in an earlier publication, \cite{dunstone08} (hereafter Paper I). This work focused on presenting the first magnetic maps of HD 155555 and the longer term evolution of magnetic and brightness features. {{A binary Zeeman Doppler imaging code was developed based upon the Doppler imaging code DoTS (\citealt{cam97dots}) and along the lines of previous binary Doppler imaging codes (e.g. \citealt{vincent93}, \citealt{hendry00} and \citealt{strassmeier03}).}} In Paper I we found that both components of HD 155555 have complex radial field magnetic topologies with mixed polarities at all latitudes. Also present were rings of azimuthal field. HD 155555 therefore appears to share properties of both young single stars and the evolved primary stars of RS CVn binaries. 

In this paper we seek further clues as to the nature of dynamo processes occurring on this young binary system. We attempt to measure the differential rotation rates of both stars. In \S \ref{sect:maps} we use the subsequent five nights to produce an additional set of surface maps that are independent of those in Paper I. These two sets of independent maps are compared using the technique of cross-correlation in \S \ref{sect:ccf}. We subsequently incorporate a latitudinal dependent shear into the binary imaging process in \S \ref{sect:diffrot} and therefore use all available data to measure the differential rotation strength. The results of our analysis are discussed in \S \ref{sect:disc} and we present our conclusions in \S \ref{sect:conc}.

\section{Observations}
\protect\label{sect:obs}

This paper is based upon spectropolarimetric observations made at the Anglo-Australian Telescope (AAT) using the University College London \'{E}chelle Spectrograph (UCLES) which was fibre fed by the SemelPol visiting polarimeter (\citealt{semel93}) mounted at the Cassegrain focus. Eleven uninterrupted nights of observations were secured on our target star HD 155555 from 2007 March 30 to April 09. Those observations obtained on the five nights March 31 - April 04 were analysed in Paper I. In this paper we combine all the observations to analyse the surface rotation properties of HD 155555. The spectral extraction and reduction of the complete dataset was the same as that done for the subset of data used in Paper I and we refer the reader to this paper for a full account of the exact process. In total we obtain a dataset of 106 Stokes I (intensity) and Stokes V (circularly polarised) spectra.

\section{Comparing independent maps}
\protect\label{sect:maps}
The 1.68 d orbital period of HD 155555 allows us to obtain full phase coverage in five nights. Given that we have a total timebase of eleven nights (March 30 - April 09), we can therefore create two entirely independent maps of both components of HD 155555. One set of brightness and magnetic maps has already been published in Paper I, corresponding to the 53 spectra taken between March 31 and April 04 (hereafter referred to as dataset/epoch one). We therefore use the subsequent five nights, April 05 to April 09 (hereafter referred to as dataset/epoch two), to create a second set of maps. Ten fewer spectra (43) are available for this second epoch which still provides excellent phase coverage. We note that the observations taken on March 30 are not used in this part of our analysis. We prefer to have two equal five night datasets and therefore to obtain each complete map in as short a time as possible (to minimise the effect of differential rotation and evolution within each map). The effective time between the mid-points of the two epochs is five nights which is almost exactly three stellar rotations.

\subsection{Brightness maps}
\protect\label{sect:spotmaps}
As in Paper I, we use the Doppler imaging code `DoTS' (\citealt{cam97dots}) to map the surface brightness distribution of the surfaces of both stars using the Stokes I intensity spectra. The process is described in detail in Paper I and references therein. {{In Paper I standard tests were performed to explore the level of cross-talk in the maps of the two stellar components. We found this to be minimal due to the fact that for most of the orbit (approximately two thirds) the profiles of the two stars are cleanly separated in velocity space. The DoTS code has been extensively tested on binary systems with more entangled spectra than HD 155555, for example the contact system AE Phe was mapped by \cite{barnes04}.}} The signal-to-noise (S/N) of this second five nights of spectra range between 95 - 210 in the peak order and are thus similar to that of the first. For brevity we do not show the obtained fits to the Stokes I data in this paper but instead refer the reader to Paper I where the fits to the first dataset can be examined. 

\begin{table}
\caption{Summary of the orbital and physical parameters for the HD 155555 system taken from Paper I and references therein.}
\protect\label{tab:syspar}
\begin{center}
\begin{tabular}{ccc}
\hline
Element & Unit & Value \\
\hline	
Orbital: \\
$P$		& (days) 	& 1.6816463			 \\
$T_{0}$		& (HJD) 	& 2446997.9102			 \\
$\Phi_{0}$	& 		& 0.752474			 \\
$e$		&  		& 0.0 		 		 \\
$\gamma$	& (\kms) 	& 3.72$\pm$0.02  		 \\
$q$	& $m_{2}/m_{1}$	& 0.935$\pm$0.001  		 \\
$K_{1}$		& (\kms) 	& 86.4$\pm$0.1	 		\\
$K_{2}$		& (\kms)	& 94.7$\pm$0.2  		 \\
$i$		& (degs)	& 50$\pm$5 (52 adopted)	\\
$(v \sin i)_1$  & (\kms)        & 34.9				 \\
$(v \sin i)_2$  & (\kms)        & 31.3				 \\
\multicolumn{2}{c}{Physical (assuming $i=52$\degs):}  & \\
$m_{1}$		& ($M_{\odot}$)	& 1.054				 \\
$m_{2}$		& ($M_{\odot}$)	& 0.986				\\
$R_{1}$		& ($R_{\odot}$)	& 1.47				\\
$R_{2}$		& ($R_{\odot}$)	& 1.32				\\
$T_{{\rm{eff}}, 1}$	& (K)		& 5300$\pm$100		\\
$T_{{\rm{eff}}, 2}$	& (K)		& 5050$\pm$100		\\
${\rm{log}}\ g_{1}$&	 	& 4.05$\pm$0.1			 \\
${\rm{log}}\ g_{2}$& 		& 4.10$\pm$0.1			 \\
\hline
\end{tabular}
\end{center}
\end{table}

\begin{figure*}
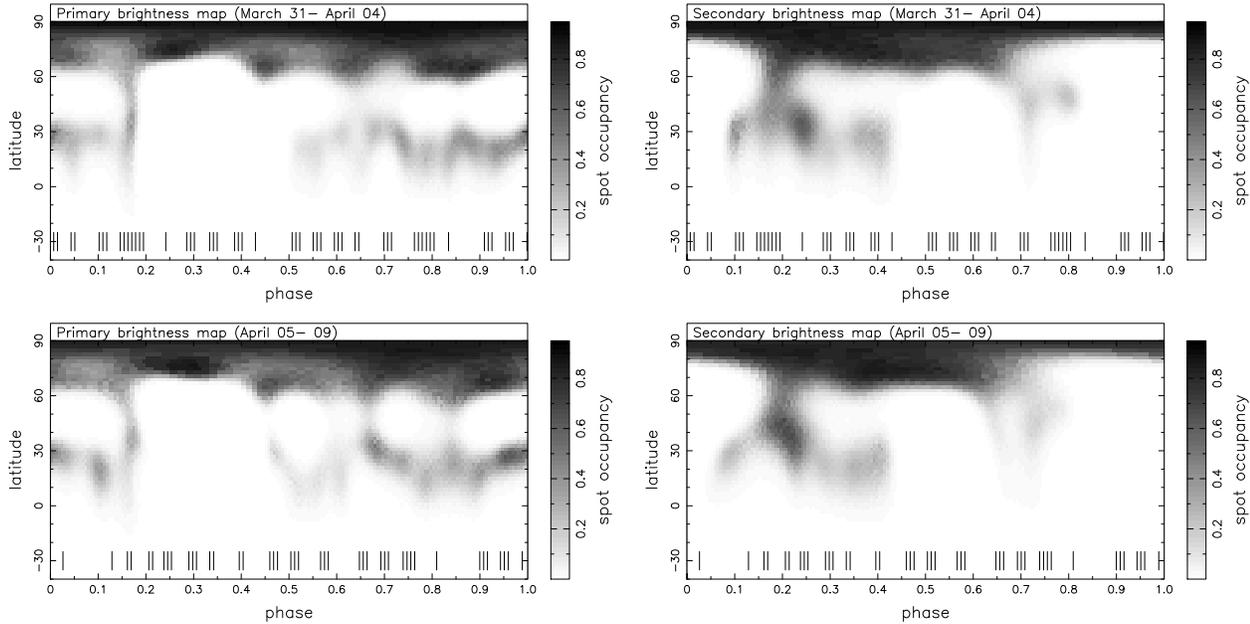

 \begin{center}
  \begin{tabular}{cc}
    \includegraphics[width=4.cm,angle=270]{Image/hd155555df_fig1.ps} &
    \includegraphics[width=4.cm,angle=270]{Image/hd155555df_fig2.ps} \\
    \includegraphics[width=4.cm,angle=270]{Image/hd155555df_fig3.ps} &
    \includegraphics[width=4.cm,angle=270]{Image/hd155555df_fig4.ps} \\
  \end{tabular}
 \end{center}
\caption[Spot maps from both epochs]{Spots maps for both stars (primary - left, secondary - right) are plotted from the original Paper I epoch (top) and the new epoch (bottom). Vertical tick marks show the phases of observation.}
\protect\label{fig:spotmaps}
\end{figure*}

Observations are phased according to the ephemeris of \cite{strass00}, see Table \ref{tab:syspar}. Also listed in Table \ref{tab:syspar} are the system parameters for HD 155555 which were obtained in Paper I {{using the \chisq\ minimisation technique of \cite{barnes00}.}} These orbital and physical parameters for the two stars are used in the Doppler imaging process. In Fig. \ref{fig:spotmaps} we present the brightness (spot) maps recovered from both epochs, for both stars.

It is clear from Fig. \ref{fig:spotmaps} that both epochs produce very similar maps for the surface brightness distribution. The features described in Paper I are present also for the second dataset. On close examination of the primary star map, a weak low latitude spot is recovered at phase $\phi = 0.45$ in the second dataset which has no counterpart in the first. The secondary star also has the major spot groups on the second epoch as were described in the first dataset. The two mid-latitude spots at phases $\phi = 0.7 - 0.8$ are not as well defined on the second dataset as the first. In summary, the maps show little evidence of major evolution between the two datasets. The small differences that we do observe may be caused by one or more of the following factors: a) the presence of differential rotation; b) flux emergence and diffusion; c) small changes caused by differences in the phase sampling of the two datasets.

\subsection{Magnetic maps}
\protect\label{sect:magmaps}

We use our new ZDI code, `ZDoTS', that was developed in Paper I (based upon the single-star ZDI code of \citealt{hussain00}), to recover the magnetic maps of HD 155555 from the Stokes V spectra. ZDoTS is capable of modelling the contribution from each star to the combined Stokes V spectra and so makes use of the entire dataset, including conjunction phases. {{As in the single star ZDI code of \cite{hussain00}, we model the local Stokes V assuming the intrinsic profile is Gaussian (which is fitted to the Stokes I profile) and constant over the stellar surface. We further assume that the weak field approximation holds. This is found to be the case for fields up to several kG from observations of very slowly rotating and weakly magnetized Ap stars (see \citealt{donati97ab}). No assumption is made about the underlying photospheric temperature. The individual weights ($w$) assigned to each spectral line in the creation of the combined LSD profile is: $w=g{\lambda}d$ (where $g$ is the magnetic Land\'{e} factor, $\lambda$ is the wavelength and $d$ is the line depth). Inverse-variance weighting is then used to recover as optimally weighted and unbiased a representation of the composite Stokes V profile as possible. We refer the reader to \cite{donati97survey} and \cite{donati97recon} for a full discussion of the implications of this approach. In addition to the tests of the ZDoTS code outlined in Paper I, further tests using synthetic data have been performed in \cite{dunstone08thesis}.}}

{{The magnetic polarity maps we recover yield only the mean magnetic flux density per surface resolution element, weighted by the mean photospheric brightness. Due to this fact the magnetic polarity signals from regions of low photospheric surface brightness (e.g. starspots) will be suppressed. The magnetic maps should therefore be considered as being representative mainly of the bright magnetic network. For the purposes of the present study this suppression has little influence on the differential rotation measurements because we are simply looking for identifiable small-scale magnetic structures that can act as tracers of the local differential rotation rate.} In Fig. \ref{fig:magmaps} we show the recovered radial field maps of HD 155555 for the two datasets. We do not show here the magnetic maps for the azimuthal field as the radial maps are sufficient to allow the reader to appreciate the temporal evolution of the magnetic structures (the reader is again referred to Paper I if interested in examining the azimuthal field maps for the first epoch).

\begin{figure*}
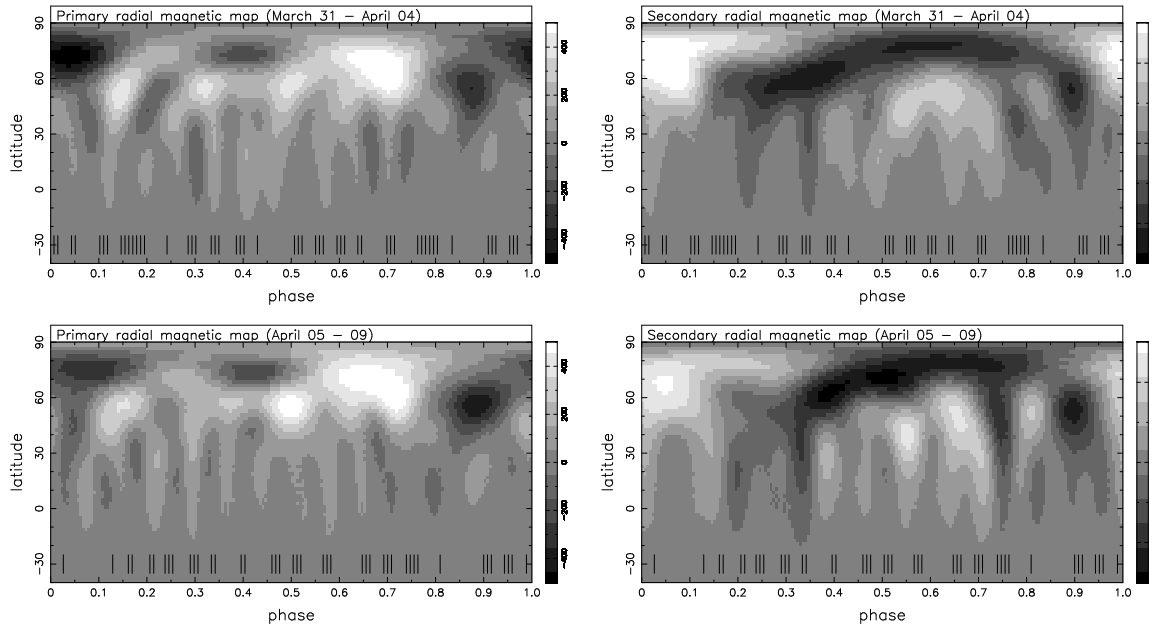

 \begin{center}
  \begin{tabular}{cc}
    \includegraphics[width=4.cm,angle=270]{Image/hd155555df_fig5.ps} &
    \includegraphics[width=4.cm,angle=270]{Image/hd155555df_fig6.ps} \\
    \includegraphics[width=4.cm,angle=270]{Image/hd155555df_fig7.ps} &
    \includegraphics[width=4.cm,angle=270]{Image/hd155555df_fig8.ps} \\
  \end{tabular}
 \end{center}
\caption[Radial magnetic maps from both epochs]{Radial magnetic field maps for both stars (primary - left, secondary - right) are plotted from the original Paper I epoch (top) and the new epoch (bottom). Vertical tick marks show the phases of observation.}
\protect\label{fig:magmaps}
\end{figure*}

As with the brightness maps, the radial magnetic maps of both components from the two epochs look very similar at first glance. On closer inspection, a number of features have changed their exact locations on the stellar surfaces, particularly on the primary star. As we shall determine in the following sections this is mainly due to the strong differential rotation for the magnetic features on the primary star. 

\section{Cross-correlating maps}
\protect\label{sect:ccf}

The first technique that was employed to measure stellar differential rotation using Doppler imaging involved cross-correlating image maps. This was done for the early differential rotation measurements on the K0 dwarf AB Dor by \cite{donati97ab} and \cite{donati99ab}. The technique requires two independent maps that have been derived from observations spanning at least two rotations of the star. This is not difficult to achieve on the AB Dor as the 0.51 d rotational period allows almost complete phase coverage in a single night. Therefore even a two night observing run could, in theory, be used for cross-correlation. However, in practise a couple of nights gap is normally left between epochs to allow enough time for measurable shear to occur. In comparison, the 1.68 d rotational period of the components of HD 155555 means that in 5 nights of observing we achieve complete phase coverage. From the last section we have two independent maps that are effectively separated by 5 nights (3 stellar rotations).

The cross-correlation (\citealt{simkin74} and \citealt{tonry79}) is performed on each latitude slice of the images. This is carried out in a way that accounts for the continuous nature of the stellar surface, i.e. that the surface maps in Figs. \ref{fig:spotmaps} \& \ref{fig:magmaps} wrap around. The resulting cross-correlation functions are shown in Fig. \ref{fig:ccf} and are examined for the characteristic signature of differential rotation. We find the peak of each latitude strip's cross-correlation function by fitting a parabola. This then allows us to obtain sub-pixel accuracy in the resulting fits. Latitudes above 70\degr\ suffer from significant longitudinal smearing due to the poorer spatial resolution at high latitudes and so are not used. Similarly our maps contain no significant information below a latitude of -30\degr. The peaks of the cross-correlation functions are then themselves fitted with a solar-like sin$^2\ l$ differential rotation law (see below in \S \ref{sect:diffrot}) which is overplotted on Fig. \ref{fig:ccf}.

\begin{figure*}
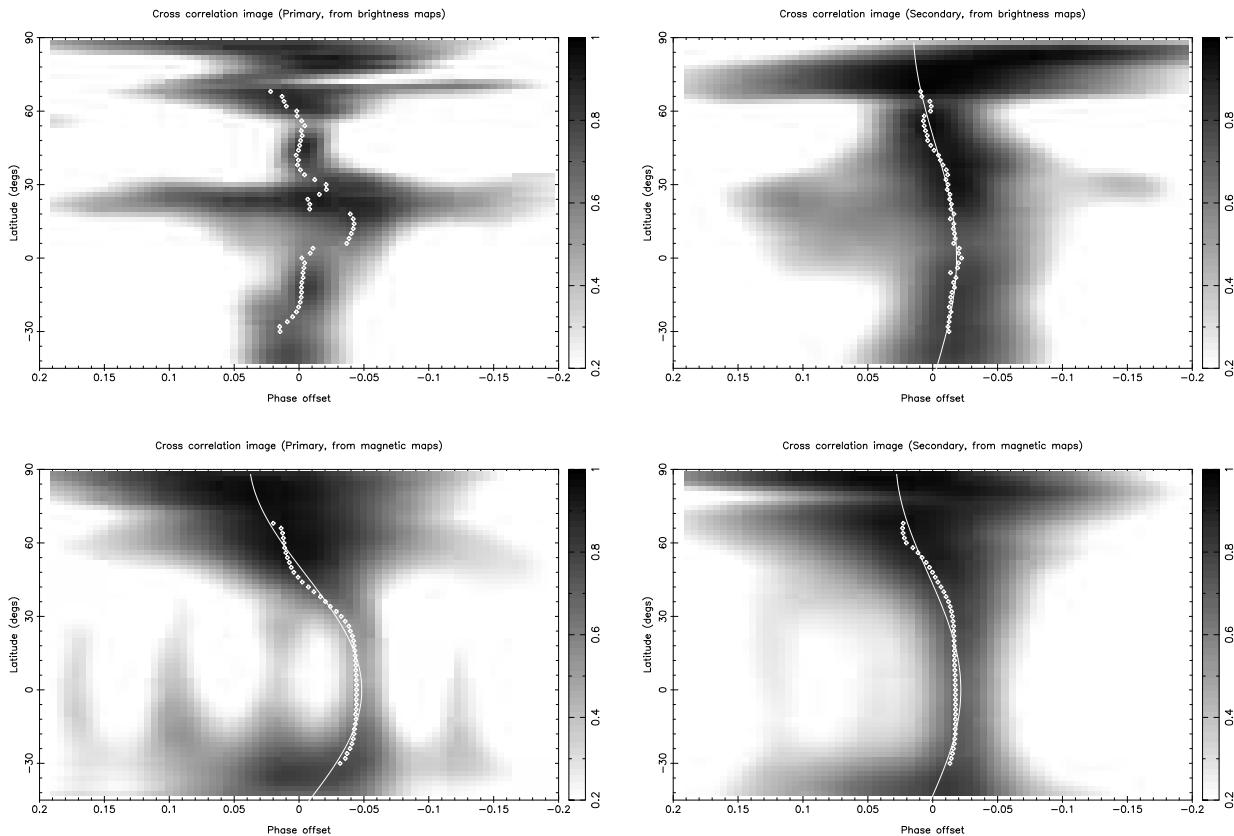

 \begin{center}
  \begin{tabular}{cc}
    \includegraphics[width=5.5cm,angle=270]{Image/hd155555df_fig9.ps} &
    \includegraphics[width=5.5cm,angle=270]{Image/hd155555df_fig10.ps} \\
    \includegraphics[width=5.5cm,angle=270]{Image/hd155555df_fig11.ps} &
    \includegraphics[width=5.5cm,angle=270]{Image/hd155555df_fig12.ps} \\
  \end{tabular}
 \end{center}
\caption[Cross-correlation images]{The independent maps from the two epochs are cross-correlated to search for latitudinal dependant surface rotation. The two panels on the left are for the primary star, while the two on the right show the secondary star. The top plots come from cross-correlating the spot maps and the bottom plots from the geometric average of the cross-correlation images of the radial and azimuthal field maps. Plus symbols mark the maximum of the cross-correlation function for each latitude strip and the solid lines are the best fits of a simple solar-like differential rotation law for all but the primary spot cross-correlation image (see text). Note that the x-axis plotting phase has been reversed so as to present the reader with the more conventional diagram of differential rotation that would have been obtained if we had adopted stellar longitude.}
\protect\label{fig:ccf}
\end{figure*}

The imaging code assumes that the rotational period of each star is equal to the binary orbital period. If this assumption of synchronous rotation is incorrect then the cross-correlation images should show an offset in phase. This would manifest itself as a vertical line offset from zero phase shift in Fig. \ref{fig:ccf}, as an incorrect period would not be latitude dependant. From an examination of Fig. \ref{fig:ccf}, it is obvious that all four maps show no evidence of uniform asynchronous rotation. The binary orbital period is the correct period at a particular stellar latitude. It is also clear that in all cases the equator is rotating faster than the poles.

Out of the four cross-correlation images in Fig. \ref{fig:ccf} the only one that we are unable to fit with a differential rotation law was that derived from the primary star's brightness images. An examination of the primary spot images in Fig. \ref{fig:spotmaps} shows why.  Spots are present near the pole (above 60\degr) and at a low latitude band between 5 and 35\degr\ but there are very few spot features at the intermediate latitudes. The corresponding primary magnetic images (Fig. \ref{fig:magmaps}) however, have structure at all latitudes and so produce a far cleaner signal as shown in the bottom-left panel of Fig. \ref{fig:ccf}. Here we find a differential rotation strength of $\Delta\Omega = 0.10\ \rm{rad\ d^{-1}}$, or an equator-pole lap time of 60 days.  If we now compare this fit with the primary brightness map we can see that at the latitudes of 5-20\degr, within the low latitude band of spots, we find evidence of a similar phase shift.

The cross-correlation of the secondary brightness map produces a much better defined differential rotation signature than the primary star. Again this is because the secondary star has spots at a larger range of latitudes. It is well fitted by a sin$^2\ l$ law and results in a differential rotation measurement of $\Delta\Omega = 0.04\ \rm{rad\ d^{-1}}$, or a equator-pole lap time of 157 days.  The corresponding cross-correlation image using the secondary magnetic maps also produces a definite signature of differential rotation with $\Delta\Omega = 0.06\ \rm{rad\ d^{-1}}$, or a near solar-like equator-pole lap time of 105 days.

\section{Image shear}
\protect\label{sect:diffrot}

Recent measurements of stellar differential rotation have been made using the sheared image technique, as developed by \cite{donati00rx}.  Instead of cross-correlating two entirely independent maps, as in the last section, a latitudinal dependant shear is applied within the imaging code to model the effect of differential rotation. As such each latitude strip has its own angular rotation rate given by:
$$\Omega(l)=\Omega_{eq}-\Delta\Omega{\rm{sin}}^{2}l$$
where $\Omega_{eq}$ is the equatorial rotation rate and $l$ is the stellar latitude. This was implemented into the DoTS code by \cite{barnes01} to work for single stars. Here we implement this into the binary version of DoTS. This is achieved by first applying the latitudinal dependant shear to each vertex of our surface pixel grid for each star and then rotating the star in the co-ordinate frame of the binary orbit. This change was also made to the our new ZDI code, ZDoTS, allowing us to obtain a separate measurement of the stellar differential rotation from the Stokes V spectra.

\begin{figure}
 \begin{center}
  \includegraphics[width=6.cm,angle=90]{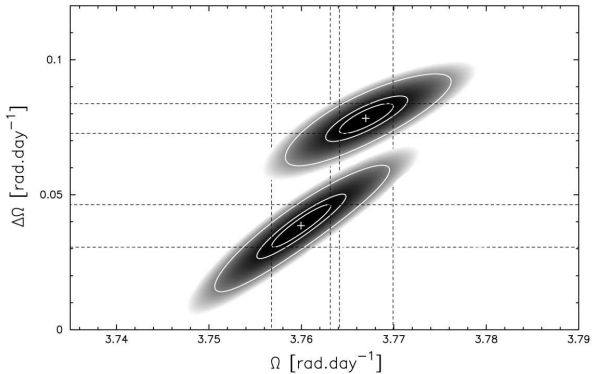}
 \end{center}
\caption[Differential rotation from image shear of spots]{The results of the image shear technique for measuring differential rotation using the spot (Stokes I) signatures. Differential rotation rate between the equator and pole is plotted as a function of absolute equatorial rotation rate. The primary star is the upper plot and is found to have stronger differential rotation than the secondary star which is the lower plot. Contours showing the one parameter 1-$\sigma$, 2.6-$\sigma$ (99\%) and 4-$\sigma$ (99.99\%) confidence intervals are superimposed over the $2{\alpha}S$ surface. }
\protect\label{fig:dfrotspot}
\end{figure}

The image shear method has many advantages over that of the original cross-correlation technique that we outlined in the last section. It works in data space directly and can therefore produce formal uncertainties on the differential rotation parameters. Many of the problems in fitting the differential rotation law to the peaks in the cross-correlation images also disappear. Furthermore, we do not require two complete and independent surface maps. Instead, all that is required is for at least one phase to be re-observed during a subsequent stellar rotation (preferably with both low and high latitude surface features). It is therefore possible to use quite incomplete datasets (poor phase coverage) and will even work for stars with pathological periods near to a multiple of a day. Also the technique is less sensitive to non-uniform latitudinal spot coverage (a problem we encountered with the brightness images of the primary star in \S \ref{sect:ccf}) as the shear is applied to the model rather than derived from the result.

The values of $\Omega_{eq}$ and $\Delta\Omega$ are found by running a grid of models which sample combinations of the two variables. We solve for each set of rotation parameters for each of the two stars separately, as there is no reason to suspect that they should be significantly correlated. When obtaining the stellar parameters (displayed in Table \ref{tab:syspar} from Paper I) we set the code to aim for a very small value of \chisq\ (in reality this is unobtainable) and then record the actual \chisq\ achieved after a fixed number of maximum entropy iterations. The best value is then found by interpolating the two dimensional grid of parameters and searching for the global \chisq\ minimum. We use a slightly different approach when obtaining the differential rotation parameters. 

The above technique does not take advantage of the image entropy information. The full relative posterior probability is $\exp(\alpha S - \frac{1}{2} \chi^{2})$, where $S$ is the entropy of the final image and $\alpha$ is the Lagrange multiplier. Therefore $2{\alpha}S$ has the same units as \chisq\ and when \chisq\ is fixed can be used for purposes of determining error regions for the fitted parameters ($\Omega_{eq}$,$\Delta\Omega$). So we choose an obtainable value of reduced \chisq\ (e.g. \chisq=0.7 for the Stokes I data) and then record the value of $2{\alpha}S$ once the code has converged. This is subtly different to the former technique and the probability contours are found to be smoother than that produced when using the minimum \chisq. This is probably due to the fact that the recovered image becomes dominated by spurious noise when pushing the image to small \chisq\ and so is more susceptible to local minima. 

We apply the sheared image technique, as described above, to the entire 11 nights of observations (including the so far unused observations on March 30). In Fig. \ref{fig:dfrotspot} we show the log posterior probability ($2{\alpha}S$) surface defined by the sets of differential rotation parameters for each star from the intensity (Stokes I spectra) along with probability contours. As the two stars have different strengths of differential rotation it has been possible to display both stars on the same plot. Similarly in Fig. \ref{fig:dfrotmag} we produce the same diagram but using the magnetic (Stokes V) spectra. The numerical results for the strength of the differential rotation ($\Delta\Omega$) and the equatorial rotation rate ($\Omega_{eq}$) are tabulated in Table \ref{tab:dfrot} along with the results from fitting the cross-correlation functions in \S \ref{sect:ccf}. The tilted ellipses of Figs. \ref{fig:dfrotspot} \& \ref{fig:dfrotmag} show that the parameters $\Delta\Omega$ and $\Omega_{eq}$ are correlated. This is to be expected due to the fact that the shear rate is most tightly constrained by mid-latitude features to which the Doppler imaging process is inherently most sensitive.

\begin{figure}
 \begin{center}
  \includegraphics[width=9cm,angle=0]{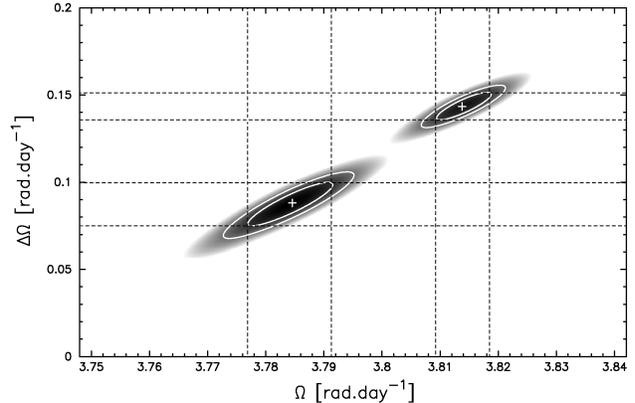}
 \end{center}
\caption[Differential rotation from image shear of magnetic features]{The results of the image shear technique for measuring differential rotation using the magnetic (Stokes V) signatures. Axis are as Fig. \ref{fig:dfrotspot}. Contours showing the one parameter 1-$\sigma$, 2.6-$\sigma$ (99\%) confidence intervals are superimposed over the $2{\alpha}S$ surface. The primary star is again found to have stronger differential rotation and is therefore the upper plot.}
\protect\label{fig:dfrotmag}
\end{figure}

\begin{table*}
\caption[Surface rotation parameters for both components of HD 155555]{Surface rotation parameters for both components of HD 155555. Columns 2 and 7, labelled `CCF', are the measurements from cross-correlating the independent maps produced from the two epochs in \S \ref{sect:ccf}.  All other columns refer to measurements and associated uncertainties of the differential rotation strength (d$\Omega$), the corresponding equator-pole lap time (`Lap'), equatorial rotation rate ($\Omega_{\rm{eq}}$) and latitude of synchronous rotation ($\theta_{s}$) produced from the sheared image technique in \S \ref{sect:diffrot}.}
\protect\label{tab:dfrot}
\begin{center}
\begin{tabular}{|l||c|c||c|c|c|c|c|c|c|c|c}
\hline
 &\multicolumn{5}{c|}{Brightness Images}& &\multicolumn{5}{c|}{Magnetic Images}\\
\cline{2-6} \cline{8-12}
         & CCF       & \multicolumn{4}{|c}{Image shear}       &    & CCF       & \multicolumn{4}{|c}{Image shear}\\
	 & d$\Omega$ & d$\Omega$   & Lap       & $\Omega_{\rm{eq}}$  & $\theta_{s}$   &    & d$\Omega$ & d$\Omega$   & Lap       & $\Omega_{\rm{eq}}$ & $\theta_{s}$\\
	 & (rad d$^{-1}$) & (rad d$^{-1}$)  & d  & (rad d$^{-1}$)  & (\degs)    &    & (rad d$^{-1}$) & (rad d$^{-1}$)  & d  & (rad d$^{-1}$) & (\degs)\\
\hline

Primary  & -         & 0.078 $\pm$ 0.006 & 80.2  & 3.767 $\pm$ 0.003 & 38.8   &   & 0.104     & 0.143 $\pm$ 0.008 & 43.8 & 3.814 $\pm$ 0.005 & 47.3 \\

Secondary& 0.040     & 0.039 $\pm$ 0.006 & 163 & 3.760 $\pm$ 0.003 & 51.7   &   & 0.060     & 0.088 $\pm$ 0.012  & 71.3 & 3.785 $\pm$ 0.007 & 47.8 \\
\hline
\end{tabular}
\end{center}
\end{table*}

The differential rotation result from the secondary star's intensity spectra of $\Delta\Omega=0.0386\pm0.0061\ \rm{rad\ d^{-1}}$ is very similar (consistent within the quoted uncertainty) to that produced from the cross-correlation technique of $\Delta\Omega=0.04\ \rm{rad\ d^{-1}}$. This could have been predicted given that the $\rm{sin}^2$ law fit to the cross-correlation function of the secondary stars spot features, in the top-right panel of Fig. \ref{fig:ccf}, was by far the most acceptable. The magnetic (Stokes V) spectra produce strengths of differential rotation consistently 30 \%\ stronger than was found from the cross-correlation technique for both the primary and secondary components. 

% This is unsurprising from looking at the magentic maps cross-correlation images in Fig. \ref{fig:ccf} (itself a product of the original maps in Fig. \ref{fig:magmaps}) which illustrates that a lack of structure in the longitudinal direction at latitudes much below about 30 \degr.

\begin{figure}
 \begin{center}
  \includegraphics[width=6cm,angle=270]{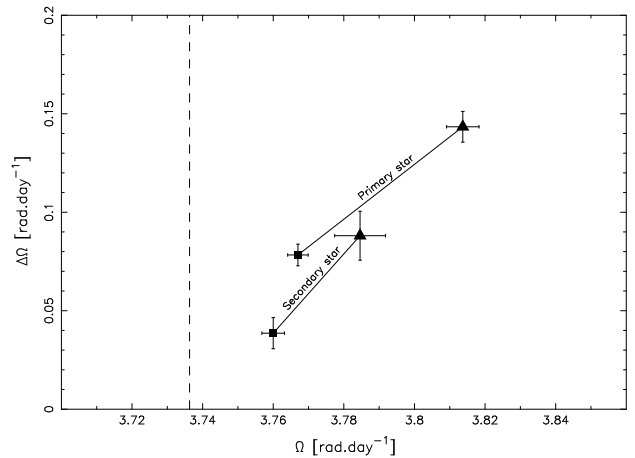}
 \end{center}
\caption[Combining differential rotation measurements]{All four differential rotation measurements are plotted with 1-$\sigma$ uncertainties. Axis are as Figs.  \ref{fig:dfrotspot} and \ref{fig:dfrotmag}.  The two squares are the measurements of differential rotation using spots and the two triangles are from magnetic features. Labelled lines join together measurements on each of the primary and secondary stars. The vertical dashed line shows the rotation rate corresponding to the binary orbital period.}
\protect\label{fig:dfrotcomb}
\end{figure}

We summarise the results of our analysis in Fig. \ref{fig:dfrotcomb} which combines all four of the differential rotation measurements from the sheared image technique. The primary star clearly has a stronger surface differential rotation rate from both the spot and magnetic features. Also the results obtained from the Stokes V spectra from both stars reveal that the differential rotation is approximately twice as strong from using the magnetic regions than the spots.

Fig. \ref{fig:dfrotcomb} also shows that the equators of both stars rotate considerably faster than the binary orbital rate (shown as the vertical dashed line in Fig. \ref{fig:dfrotcomb}). This then raises the question of whether any stellar latitude is synchronously rotating with the orbit. In order to show this graphically we plot in Fig. \ref{fig:laws} the differential rotation curves obtained for the two stars, for both the spots and magnetic signatures. We clearly find that the poles are rotating sub-synchronously while the equators rotate super-synchronously in all cases. The exact synchronously rotating latitude for each differential rotation measurement is given in Table \ref{tab:dfrot}.

\begin{figure*}
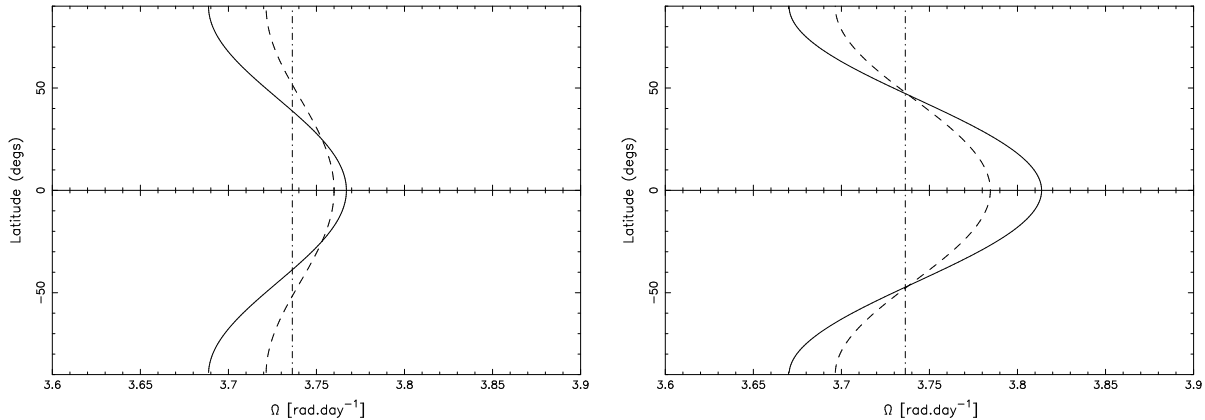

 \begin{center}
  \begin{tabular}{cc}
    \includegraphics[width=5.5cm,angle=270]{Image/hd155555df_fig16.ps} &
    \includegraphics[width=5.5cm,angle=270]{Image/hd155555df_fig17.ps} \\
  \end{tabular}
 \end{center}
\caption[Differential rotation curves]{The differential rotation curves are plotted from the spots (left) and magnetic regions (right). The solid line is the primary star and the dashed line is the secondary star. The dot-dash vertical line corresponds to the orbital rotation period and thus shows that both stars are synchronously rotating with the binary orbit.}
\protect\label{fig:laws}
\end{figure*}

Curiously the primary and secondary stars share the same synchronously rotating latitude (47\degr) for the magnetic differential rotation. The resulting synchronously rotating latitude for the primary star's spot distribution is 39\degr\ and is thus the lowest of all four measurements. This may be interesting given that this latitude marks the upper boundary of the lower latitude band of spots seen on the primary star.

\section{Discussion}
\protect\label{sect:disc}

We have found that both components of HD 155555 have non-negligible surface differential rotation. This result was obtained from using the cross-correlation and the image shear techniques. There are two reasons why the cross-correlation technique (\S \ref{sect:ccf}) should probably be carried out whenever sufficient data of a star are available. Firstly, it provides a very literal interpretation of the data and so was useful in validating our implementation of the image shear technique to binary systems. Secondly, the cross-correlation technique does not assume {\it a priori} knowledge for the form of the differential rotation law which the image shear technique does. Beyond this however, the image shear technique is superior and has many advantages which were outlined in \S \ref{sect:diffrot} and will be touched on again in \S \ref{sect:finmaps}. We note that the direct spot tracking technique of \cite{cam02} is another alternative which makes few assumptions but requires very good phase coverage and high signal-to-noise data.

\subsection{Internal velocity fields}
\protect\label{sect:intvel}

To summarise the results of \S \ref{sect:diffrot}; the primary star has stronger differential rotation than the secondary star and the magnetic regions produce stronger signatures than do the spots. The latter trend was first revealed by \cite{donati03b} for AB Dor and HR 1099. The authors suggested that the spots and magnetic regions are anchored at different depths in the convection zones of these stars. The fact that the strength of the shear is so much greater in the magnetic regions suggests that the dynamos of these stars may not be active just at the interface layer between the radiative core and the convective envelope but instead be distributed throughout the convection zone. Recent numerical simulations by \cite{brown07} lend support to this idea by showing that global-scale toroidal and poloidal fields can be generated and maintained in the convection zones of rapidly rotating stars.

\cite{donati03b} attempted to quantify the nature of the internal velocity fields in the rapidly rotating single stars AB Dor and LQ Hya by examining the temporal fluctuations in the differential rotation parameters ($\Omega_{eq}$,$\Delta\Omega$). They found that:
$$\Omega_{eq} = \lambda\Delta\Omega + \Omega_{sb}$$
where $\lambda$ depends on the assumed internal rotation model and on the internal stellar structure and $\Omega_{sb}$ is a constant, equal to the rotation rate the stellar convection zone would have if spinning as a solid body. 

For HD 155555 we only have a single epoch of differential rotation measurements however, given the binary nature of the system, we do have a very useful zero-point. We know that in the absence of differential rotation, the rotation rate of each star would be equal to that of the orbital rotation rate. Therefore we can measure $\lambda$ directly from Fig. \ref{fig:dfrotcomb}. On this plot the gradient of the lines joining the zero point (where $\Omega_{eq}=\Omega_{sb}$ and $\Delta\Omega=0$) with each of our differential rotation measurements will be equal to $\lambda^{-1}$. We obtain gradients in the range 1.6 to 2.5. \cite{donati03b} considered two possible models for the angular velocity fields. Firstly, that angular rotation is constant with radius (like it is for the Sun) then our gradient ($\lambda^{-1}$) should be equal to five. Secondly that angular velocity was constant along cylinders symmetric to the rotation axis, resulting in a gradient of approximately two (for stars such as AB Dor and the components of HD 155555). Therefore, like AB Dor, our results show that a model of internal rotation being constant along cylinders is most appropriate for the components of HD 155555. Given their relatively rapid rotation this is what one would have expected.

By adopting this model of internal rotation, we can follow the example of \cite{donati03b} and calculate the range of possible latitudinal shears. At a given latitude the rotation rate will depend upon how deep the tracers of differential rotation are anchored in the stellar convection zone. The total depth of the convection zone of each star can be expressed as the fractional radius at the base of the convection zone, $r_{c}/R$ (where a star with $r_{c}/R$=0 is fully convective and a star with $r_{c}/R$=1 has no convection zone). Then the maximum expected ratio between the shears we measure from the Stokes V and I data should be 1/$(r_{c}/R)^2$. So if the observed value were to equal this maximum ratio then this would suggest that the magnetic regions are anchored near to the stellar surface while the spot features were anchored near the base of the convection zone. The value of $r_{c}/R$ can be obtained using the evolutionary models of \cite{siess00}. The components of HD 155555 have $r_{c}/R=0.68$ for the primary and $r_{c}/R=0.63$ for the secondary star. This corresponds to a maximum ratio of 2.16 for the primary star and 2.52 for the secondary star. Using our results for HD 155555, we find that the observed ratio between the shears from Stokes V and I is 1.8 for the primary star and 2.25 for the secondary star. We therefore find that HD 155555 is consistent with the adopted model of internal rotation.

\subsection{Comparison with other systems}

\begin{figure*}
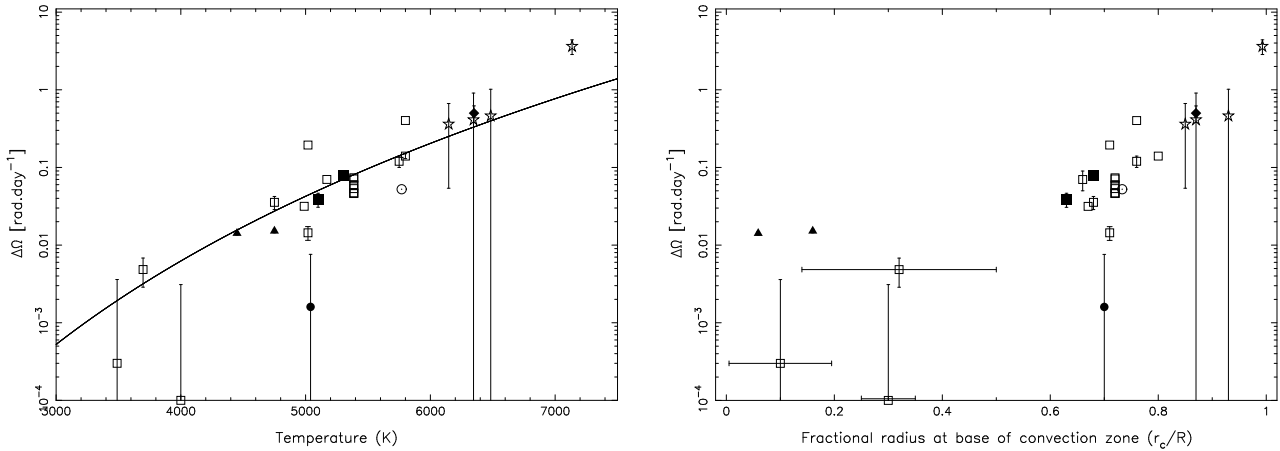

 \begin{center}
  \begin{tabular}{cc}
    \includegraphics[width=6.cm,angle=270]{Image/hd155555df_fig18.ps} &
    \includegraphics[width=6.cm,angle=270]{Image/hd155555df_fig19.ps} \\
  \end{tabular}
 \end{center}
\caption[Differential rotation is plotted as a function both stellar temperature and convection zone depth]{Left: Differential rotation is plotted as a function of stellar temperature. The solid line shows the power law fit of \cite{cameron08}. Right: The x-axis now plots the fractional radius at the base of the convection zone ($r_{c}/R$) as estimated from stellar evolutionary models (see text). Both plots include the same stars; filled symbols are members of binary systems, open symbols are single stars. The two triangles are the evolved RS CVn systems HR 1099 and IM Peg. The filled circle is the pre-cataclysmic system V471 Tau. Open squares are single G, K and M dwarfs, the `star' symbols are F stars and the circled dot represents the Sun. The two large filled squares are the components of HD 155555. Error bars are plotted when the uncertainty significantly exceeds the symbol size.}
\protect\label{fig:dfrotjrb}
\end{figure*}

It is now possible to compare our results with those of both single rapidly rotating stars and binary systems. In the left panel of Fig. \ref{fig:dfrotjrb} we plot differential rotation rate versus effective temperature. A similar plot was first produced for single G, K and M dwarfs by \cite{barnes05}. The authors discovered that for single stars the strength of differential rotation increases with increasing stellar surface temperature. More recently \cite{reiners06} combined these results with differential rotation measurements on F dwarfs derived from a Fourier analysis of the rotational broadening profile. An extrapolation of the \cite{barnes05} power-law fit showed that the two techniques were in good agreement. The latest version of this diagram was made by \cite{cameron08} who performed a re-analysis of the power law fit using all latest available differential rotation measurements. This is plotted as the solid line on Fig. \ref{fig:dfrotjrb}. When we plot the differential rotation parameters for both components of HD 155555 we find that they effectively straddle this line. It therefore appears that the strength of differential rotation on both stars is consistent with single stars of the same temperature.

In addition to the single stars, collated in \cite{barnes05} and \cite{cameron08}, we also consider the results from previous Doppler imaging studies of binary star systems.  Fig. \ref{fig:dfrotjrb} plots (as solid triangles) the differential rotation measurements of the evolved primary stars of the RS CVn binaries HR 1099 and IM Peg (from \citealt{petit04} and \citealt{marsden07} respectively). The main-sequence component of the pre-cataclysmic variable star V471 Tau (\citealt{hussain06}) is also plotted as a filled circle on Fig. \ref{fig:dfrotjrb}. All three systems show weak differential rotation ($\Delta\Omega<0.02\ \rm{rad\ d^{-1}}$). We note however that there is only a single epoch measurement for V471 Tau and this is consistent with zero ($\Delta\Omega=0.0016\pm0.006\ \rm{rad\ d^{-1}}$) differential rotation. The recovered Doppler images suffer from a lack of spot coverage at all latitudes which may make them unreliable. We therefore encourage future observations to confirm this measurement, but we do not consider V471 Tau further here.

\subsection{Tidal forces or internal structure?}

The RS CVn systems HR1099 and IM Peg have been repeatedly observed, so many measurements of differential rotation are available. When compared to single main-sequence stars of similar mass HR 1099 and IM Peg have greatly reduced differential rotation. For example, HR 1099 has a mass of 1 \msun\ and yet exhibits a differential rotation rate approximately a third that of the Sun (or approximately an eighth that of the power-law fit shown on Fig. \ref{fig:dfrotjrb}). \cite{petit04} suggested that the reduced differential rotation was due to binary tidal forces. Over time these have acted to enforce synchronous rotation at all latitudes (i.e. to inhibit differential rotation).  The relatively strong differential rotation that we find on both components of the young HD 155555 system adds further complexity to the argument. We have shown that the mere fact that a star is a member of a close binary system does not necessarily result in a reduced differential rotation rate. We are therefore left with two possibilities; that the timescale for tidal forces to suppress the differential rotation is long compared to the age of the HD 155555 system, or that binarity is unimportant and evolutionary state is the dominant factor. We now examine the first of these possibilities.

%turbulent viscosity

Our results for the young (18 Myr, \citealt{strass00}) HD 155555 system show that tidal forces have indeed enforced synchronous rotation of both components, as shown in Fig. \ref{fig:laws}. However, they have been unable to suppress the differential rotation, i.e. tides have not yet been able to synchronise rotation at all latitudes of the convective zone. Both components of HD 155555 are still contracting on to the main sequence. One may naively think that this would result in the stars rapidly spinning up and so might explain why the tidal forces are unable to effectively inhibit differential rotation. Given that HD 155555 is a close binary system we should examine the efficiency of tidal forces. The synchronisation timescale ($t_{sync}$) can be approximated as $t_{sync}{\simeq}q^{-2}(a/R)^{6}$ (\citealt{zahn77}), where $q$ is the mass ratio. Using the parameters for HD 155555, we find a timescale of order $t_{sync}\simeq10^{4}$ yrs for both components. Given that this is very much shorter than evolutionary timescales, it appears that tidal forces have ample opportunity to suppress differential rotation. Indeed, as the stars try to spin up, they will rapidly become resynchronised, with angular momentum being transferred from the stars' rotation to the binary orbit. The net result is that the stars will spin down rather than up. The fact that differential rotation is apparently uninhibited on the components of HD 155555 then places doubt on this mechanism also being responsible for the low differential rotation observed on the evolved HR 1099 and IM Peg systems. We futher note that theoretical work (\citealt{scharlemann82}) predicts that tides should not inhibit differential rotation on RS CVn systems.

If binarity is unimportant then the strength of differential rotation is governed primarily by the stars evolutionary stage. HR 1099 and IM Peg are both evolved giants and so have large radii (3.7 and 13.3 \rsun\ respectively) and correspondingly cool surface temperatures. Therefore when we plot HR 1099 and IM Peg on the left panel of Fig. \ref{fig:dfrotjrb} (solid triangles) we find that they show strengths of differential rotation very close to (possibly slightly below but within the scatter) that of a main sequence single star with a similar surface temperature. This would seem to be a major clue as to the source of the low differential rotation. If tidal forces were solely responsible then there would be no reason to expect such an agreement with single stars. These giant stars have deep convection zones, like those of the cool K and M dwarfs, and therefore their internal structure may determine their differential rotation rates. In an attempt to quantify this we use the evolutionary models of \cite{siess00} which quote the fractional radius at the base of the convection ($r_{c}/R$). This constant was discussed earlier in \S \ref{sect:intvel}.

We use the evolutionary models to estimate the value of $r_{c}/R$ for all stars that were plotted on the temperature plot of Fig. \ref{fig:dfrotjrb}. We then plot differential rotation as a function of $r_{c}/R$ in the right-hand panel of Fig. \ref{fig:dfrotjrb}. Where the models of \cite{siess00} did not follow the stellar evolution far enough the models of \cite{claret04} were used. The models are spaced in 0.1 \msun\ intervals and so we use the nearest one appropriate to each stars mass which is obtained from the literature. We only use F-stars from \cite{reiners06} that were defined as cluster members so that the cluster age could be used as a constraint to determine the stars position in the evolutionary models. The single M dwarfs are particularly difficult to obtain values for $r_{c}/R$. This is because their evolutionary timescales are longer and so the depth of the convection zone continues to change for a longer time. Therefore we show approximate ranges for the value of $r_{c}/R$ where appropriate. 

In Fig. \ref{fig:dfrotjrb} we cover the entire range of convection zone depths. A strong trend of increasing differential rotation strength with decreasing convection zone depth is found. This is very similar to the surface temperature plot which is not surprising as for main-sequence objects temperature essentially defines the depth of the convection zone. The position of the components of HD 155555 therefore match well with the other single main-sequence stars. The RS CVn binaries HR 1099 and IM Peg move over to the left of the plot with values of $r_{c}/R=0.16$ and $r_{c}/R=0.059$ respectively. This illustrates their huge convection zone depths. At a specific temperature the convective depth is larger for an evolved star than for a main-sequence star. This would certainly seem to explain the observed weak differential rotation. In fact HR 1099 and IM Peg have stronger differential rotation than would be expected for main sequence stars of similar fractional convection zone depth.

We note that for consistency we have only considered the differential rotation results from stellar brightness maps (Doppler imaging). As found in \S \ref{sect:diffrot} the shear strength is often higher for magnetic regions than for star spots. Previous studies have shown that it is often easier to measure differential rotation using magnetic maps because there is normally structure present at all stellar latitudes. It would therefore be interesting to extend the scope of this analysis and consider results from ZDI also. For example, \cite{morin08} found very weak differential rotation ($\Delta\Omega=0.0063\pm0.0004$) on the fully convective M4 dwarf V374 Peg. In another recent study \cite{donati08} found that the planetary host star $\tau$ Boo has very strong differential rotation ($\Delta\Omega=0.50\pm0.12$). This measurement is plotted on Fig. \ref{fig:dfrotjrb} (solid diamond) and has a position appropriate for its F7V spectral type. This is despite the fact that the stellar surface has apparently been forced to rotate synchronously with the planetary orbital period (\citealt{catala07}). So $\tau$ Boo lends further support to the argument that tidal forces are unable to suppress differential rotation.

\subsection{Final maps}
\protect\label{sect:finmaps}
The differential rotational parameters for both stars can now be incorporated into the imaging process.  We use the entire 11 night (106 spectra) dataset with the respective differential rotation parameters for spots and magnetic features on each star to produce the brightness and radial magnetic maps shown in Figs. \ref{fig:finspot} and \ref{fig:finmag}. We note that the \chisq\ achieved for these reconstructed images (\chisq=0.7 for Stokes I and \chisq=1.0 for Stokes V) are essentially the same as the \chisq\ for the individual images presented in \S \ref{sect:maps}. This would indicate that there has been little evolution in the spots or magnetic regions (other than the effect of differential rotation) over the eleven days. In order to show the effect of the differential rotation we also plot the primary radial field map obtained if we do not apply the differential rotation law (top panel of Fig. \ref{fig:finmag}). By including the effect of differential rotation the middle panel of Fig. \ref{fig:finmag} shows a higher level of detail. This is especially true at low latitudes where the effects of differential rotation are most noticeable. 

\begin{figure*}
\begin{center}
  \begin{tabular}{c}
    \includegraphics[width=6.0cm,angle=270]{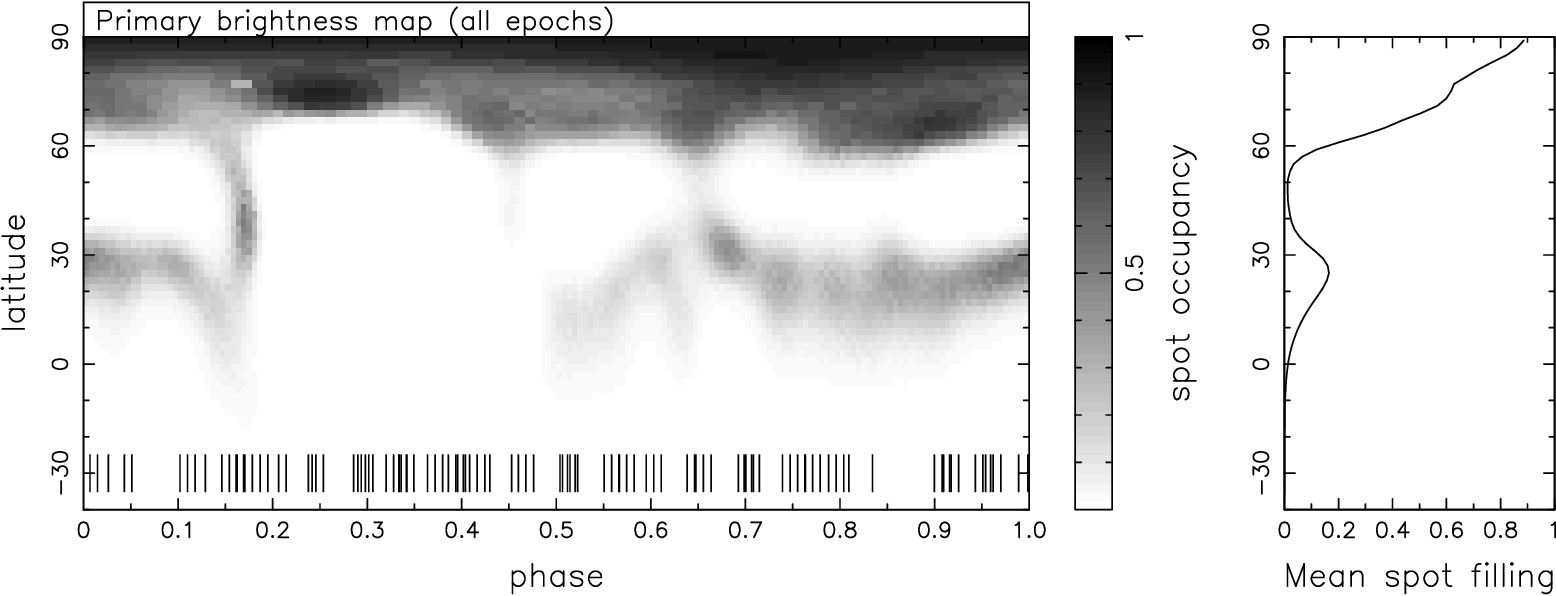} \\
    \includegraphics[width=6.0cm,angle=270]{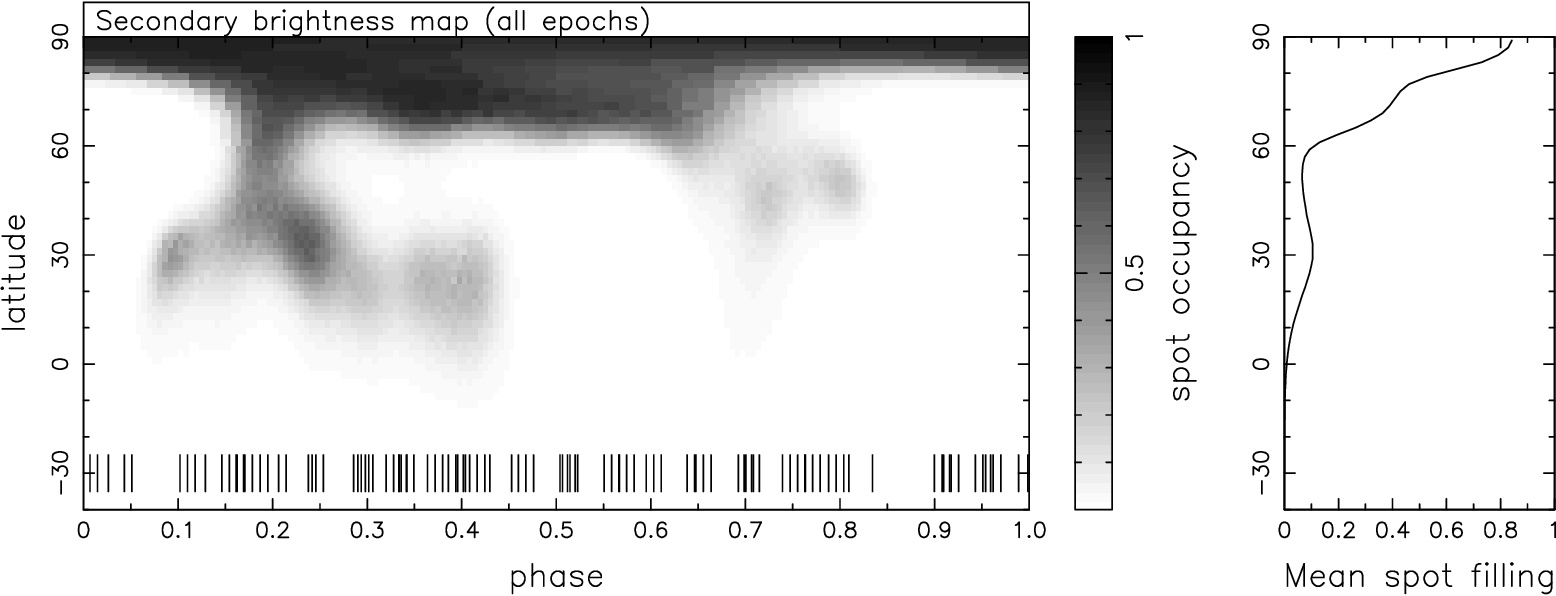} \\
  \end{tabular}
\end{center}
\caption[Final spot maps]{Final brightness maps of the primary and secondary stars produced from all eleven nights and using the obtained values of differential rotation.}
\protect\label{fig:finspot}
\end{figure*}

\begin{figure*}
\begin{center}
  \begin{tabular}{c}
    \includegraphics[width=6.0cm,angle=270]{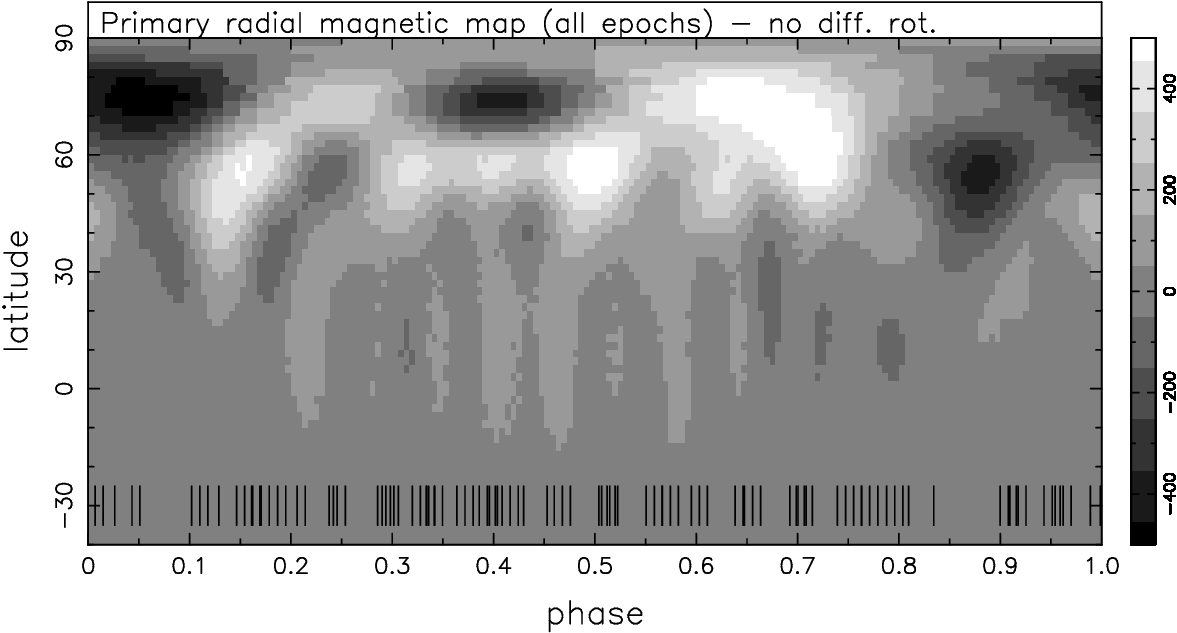} \\
    \includegraphics[width=6.0cm,angle=270]{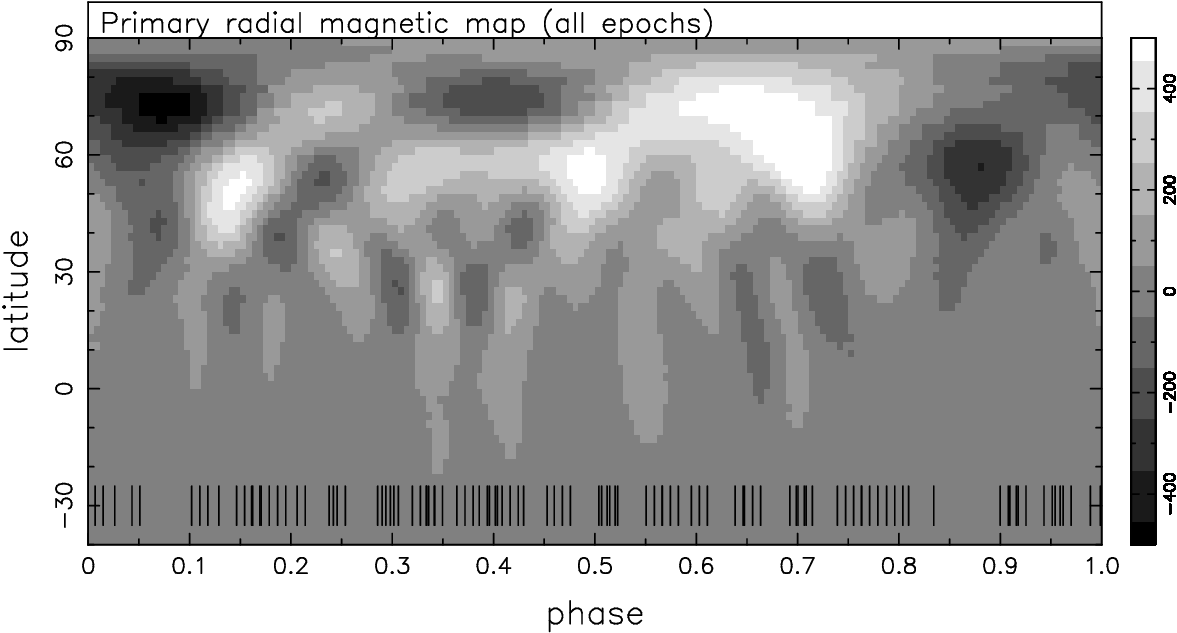} \\
    \includegraphics[width=6.0cm,angle=270]{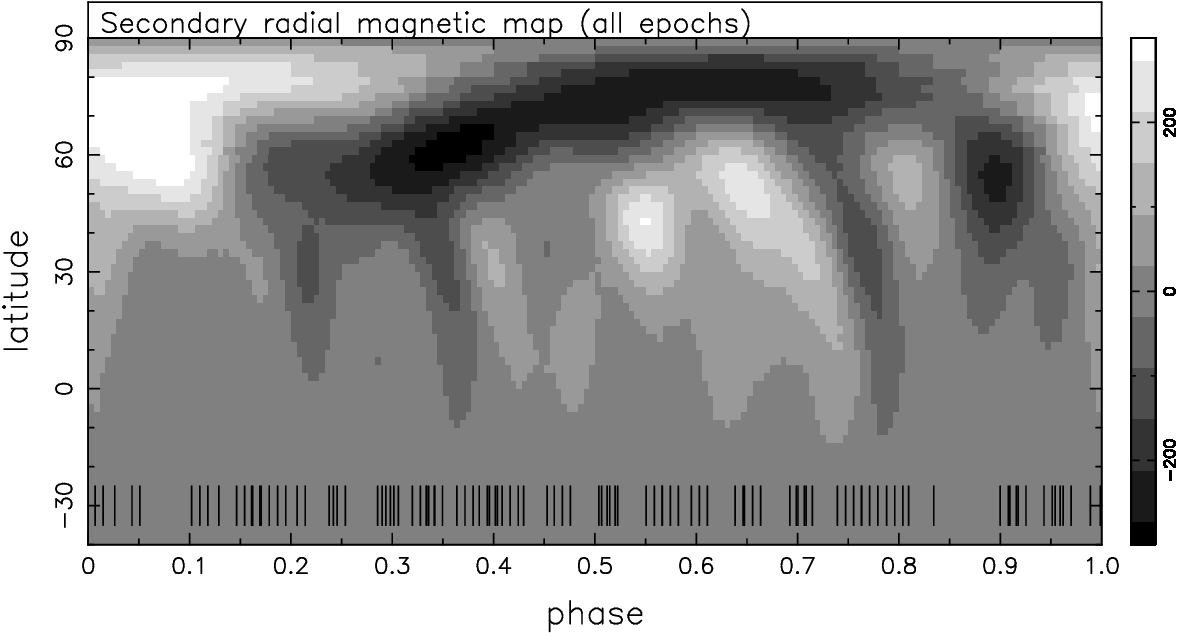} \\    
  \end{tabular}
\end{center}
\caption[Final mag maps]{Final radial field magnetic maps. The top panel shows the primary radial field map obtained from all eleven nights of observations but without incorporating the measured differential rotation into the imaging process. This should then be compared with the middle panel where the differential rotation was included, note the higher level of detail at low latitudes. The bottom panel shows the radial field map for the secondary star when using the differential rotation parameters.}
\protect\label{fig:finmag}
\end{figure*}

Looking back to Fig. \ref{fig:magmaps}, which shows magnetic maps produced from each five night subset of the data, we find that they contain more low latitude information than the top panel of Fig. \ref{fig:finmag} (no differential rotation) but less than the middle panel Fig. \ref{fig:finmag} (with differential rotation). This is what we would expect because the effect of differential rotation on only five nights is less than over eleven. Following the same argument we can now explain why the cross-correlation functions (shown in Fig. \ref{fig:ccf}) of the magnetic maps gave poor fits at low latitudes. Due to the relatively strong differential rotation, even five nights is sufficient to smear the equator of the primary star by $\Delta\phi=0.11$ and that of the secondary by $\Delta\phi=0.07$. Given that these are both larger than a whole resolution element (approximately $\Delta\phi=0.06$) it is unsurprising that the cross-correlation function fails to follow the $\rm{sin}^2$ law at low latitudes. This once again illustrates the advantage of using the sheared image technique.

Despite the improvement in the reconstruction of low latitude features, the maps presented in Fig. \ref{fig:finmag} do not change the conclusions of Paper I. It may be interesting to note however, that the small regions of flux, now well reconstructed between phases $\phi=0.3-0.5$ and latitudes 20 and 30\degs, are in the same location as the gap in the ring of spots seen for primary star in Fig. \ref{fig:finspot}. This apparent anti-correlation between the strengths of small magnetic regions and spot features may be an example of how the cooler spot temperatures suppress the contribution to the Stokes V spectra.

\subsection{Consequences for connecting field lines}

In Paper I we discussed the likelihood of field lines from one star connecting with the other. It was found from extrapolating the coronal field of each star separately (using the radial field maps) that the field complexity of the surface radial maps persists out to a significant fraction of the binary separation (7.5 \rsun or $\simeq5.5\ R_{*}$, see Table \ref{tab:syspar}) and so the interaction between the two fields is likely to be complex also. Now that we know the strength of the differential rotation on both stars, it is possible to speculate on the upper limit of the lifetimes of such binary field lines and so the energy released from the rate of re-connection of the field. Let us consider a field line that is connected from a point on the equator of the facing hemisphere of one star to another on the equator of the facing hemisphere of the other star, a situation as sketched in Fig. \ref{fig:magint}. Given that the two stars will be rotating in the same direction, but 180\degr\ out of phase, the two equatorial points will start to move away from each other due to differential rotation. The point on the primary star will sweep out a greater angle in the same time due to the stronger differential rotation (as shown in Fig. \ref{fig:magint}). 

\begin{figure}
 \begin{center}
  \includegraphics[width=8cm,angle=0]{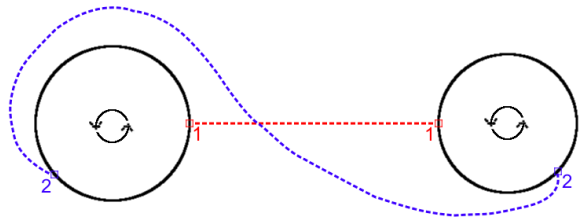}
 \end{center}
\caption[Magnetic field interaction timescales]{A cartoon sketch to show the effect of differential rotation on connecting field lines. The two components (primary - left, secondary - right) of HD 155555 and their separation are plotted to scale. A field line connects the equators of both stars. Initially at `1' the field footpoints are directly facing each other. At `2' we show the location of the footpoints 50 days later when the combined angle swept out by the two footprints is $2\pi$.}
\protect\label{fig:magint}
\end{figure}

An interesting timescale often quoted for single stars is the equator-pole laptime. On this timescale a field line that connects the pole to the equator will have become wound around the star by a full circle ($2\pi$ rads) and so stores significant magnetic energy. Similarly we can also ask; how long would it take the field points of our connecting field line to sweep out a combined angle of $2\pi$ rads? The answer is 50 days (illustrated in Fig. \ref{fig:magint}) which we note is comparable to the equator-pole lap-times of the primary star (44 days) and the secondary star (71 days). Just like the equator-pole laptimes on a single star, the equator-equator field lines connecting the two stars gives us the maximum amount of stress on the field lines. We also note however that in the absence of any force confining such field lines to the equatorial plane the connecting loop will ``balloon'' up and over the stellar poles above a critical angle (\citealt{lyndenbell94}). This would further increase the rate of field reconnection events.

The strong differential rotation provides extra stress on the field that would not be present on evolved binaries such as HR 1099 and IM Peg with their weak differential rotation. We can therefore speculate that the reconnection of these long binary field loops could significantly contribute to the X-ray luminosity of the HD 155555 system and also to the frequency of large flares. The scenario outlined above caused by differential rotation is analogous to the mechanism thought to produce the so called `superflares' on classical T Tauri stars. In this case field lines that connect the star with its surrounding disk become stretched due to the fact that the disk is rotating slower than the star (e.g. \citealt{montmerle00}, Fig. 2).

\section{Conclusions}
\protect\label{sect:conc}

We have measured the surface rotation properties of both stars in the HD 155555 system. A cross-correlation of the latitude strips of both brightness and magnetic images, produced from two independent datasets, revealed the characteristic signature of differential rotation and confirmed that this can be modelled by a simple solar-like $\rm{sin}^{2}$ law. In order to better quantify the strengths of the differential rotation we adapted the sheared image technique to binary systems. The rotational periods of both stars are found to be synchronously rotating with the binary orbital period. This confirms the fact that the system is tidally locked.

Considering only the results from the brightness maps of the surface spot distributions, the primary has an equator-pole lap time of $80$ d and the secondary star $162$ d. These results show that the strength of differential rotation on both stars is not only non-negligible but are of similar magnitude to that found on single rapidly rotating stars of the same effective temperature. We used these findings to explore the relative effects of binarity and evolutionary stage on the strength of differential rotation. The large convection zone depths of the evolved giants in RS CVn systems is more likely to be the cause of their low differential rotation rates than the effects of tidal forces. 

Future measurements of differential rotation on the components HD 155555 are strongly encouraged. It would be very interesting to examine whether temporal changes in the differential rotation rate of the components of HD 155555 are correlated. This may indicate evidence of a joint activity cycle.

%SOMETHING TO TIE BACK IN WITH PAPER I, I.E. THAT THE FIELD LOOKED SO DIFFERENT IS LINKED WITH THE DIFF ROT BEEN SO STRONG.

\section{ACKNOWLEDGEMENTS}

We would like to thank the staff at the Anglo-Australian Telescope for their support. NJD acknowledges the financial support of a UK STFC studentship.  

\bibliographystyle{mn2e}
\bibliography{iau_journals,master,ownrefs,njd2}

\begin{thebibliography}{}

\bibitem[\protect\citeauthoryear{{Barnes}}{{Barnes}}{2005}]{barnes05}
{Barnes} J.~R.,  2005, MNRAS, 364, 137

\bibitem[\protect\citeauthoryear{{Barnes}, {Collier Cameron}, {James} \&
  {Donati}}{{Barnes} et~al.}{2000}]{barnes00}
{Barnes} J.~R.,  {Collier Cameron} A.,  {James} D.~J.,    {Donati} J.-F.,
  2000, MNRAS, 314, 162

\bibitem[\protect\citeauthoryear{{Barnes}, {Collier Cameron}, {James} \&
  {Donati}}{{Barnes} et~al.}{2001}]{barnes01}
{Barnes} J.~R.,  {Collier Cameron} A.,  {James} D.~J.,    {Donati} J.-F.,
  2001, MNRAS, 324, 231

\bibitem[\protect\citeauthoryear{{Barnes}, {Lister}, {Hilditch} \& {Collier
  Cameron}}{{Barnes} et~al.}{2004}]{barnes04}
{Barnes} J.~R.,  {Lister} T.~A.,  {Hilditch} R.~W.,    {Collier Cameron} A.,
  2004, MNRAS, 348, 1321

\bibitem[\protect\citeauthoryear{{Brown}, {Browning}, {Brun}, {Miesch},
  {Nelson} \& {Toomre}}{{Brown} et~al.}{2007}]{brown07}
{Brown} B.~P.,  {Browning} M.~K.,  {Brun} A.~S.,  {Miesch} M.~S.,  {Nelson}
  N.~J.,    {Toomre} J.,  2007, in American Institute of Physics Conference
  Series Vol.~948 of American Institute of Physics Conference Series, {Strong
  Dynamo Action in Rapidly Rotating Suns}.
pp 271--278

\bibitem[\protect\citeauthoryear{{Cameron}}{{Cameron}}{2008}]{cameron08}
{Cameron} A.~C.,  2008, Astron.~Nachr., p.~1030

\bibitem[\protect\citeauthoryear{{Catala}, {Donati}, {Shkolnik}, {Bohlender} \&
  {Alecian}}{{Catala} et~al.}{2007}]{catala07}
{Catala} C.,  {Donati} J.-F.,  {Shkolnik} E.,  {Bohlender} D.,    {Alecian} E.,
   2007, MNRAS, 374, L42

\bibitem[\protect\citeauthoryear{{Claret}}{{Claret}}{2004}]{claret04}
{Claret} A.,  2004, A\&A, 424, 919

\bibitem[\protect\citeauthoryear{{Collier Cameron}}{{Collier
  Cameron}}{1997}]{cam97dots}
{Collier Cameron} A.,  1997, MNRAS, 287, 556

\bibitem[\protect\citeauthoryear{{Collier Cameron}, {Donati} \&
  {Semel}}{{Collier Cameron} et~al.}{2002}]{cam02}
{Collier Cameron} A.,  {Donati} J.-F.,    {Semel} M.,  2002, MNRAS, 330, 699

\bibitem[\protect\citeauthoryear{{Donati} \& {Brown}}{{Donati} \&
  {Brown}}{1997}]{donati97recon}
{Donati} J.-F.,  {Brown} S.~F.,  1997, A\&A, 326, 1135

\bibitem[\protect\citeauthoryear{{Donati} \& {Collier Cameron}}{{Donati} \&
  {Collier Cameron}}{1997}]{donati97ab}
{Donati} J.-F.,  {Collier Cameron} A.,  1997, MNRAS, 291, 1

\bibitem[\protect\citeauthoryear{{Donati}, {Collier Cameron}, {Hussain} \&
  {Semel}}{{Donati} et~al.}{1999}]{donati99ab}
{Donati} J.-F.,  {Collier Cameron} A.,  {Hussain} G.~A.~J.,    {Semel} M.,
  1999, MNRAS, 302, 437

\bibitem[\protect\citeauthoryear{{Donati}, {Collier Cameron} \&
  {Petit}}{{Donati} et~al.}{2003}]{donati03b}
{Donati} J.-F.,  {Collier Cameron} A.,    {Petit} P.,  2003, MNRAS, 345, 1187

\bibitem[\protect\citeauthoryear{{Donati}, {Mengel}, {Carter}, {Marsden},
  {Collier Cameron} \& {Wichmann}}{{Donati} et~al.}{2000}]{donati00rx}
{Donati} J.-F.,  {Mengel} M.,  {Carter} B.~D.,  {Marsden} S.,  {Collier
  Cameron} A.,    {Wichmann} R.,  2000, MNRAS, 316, 699

\bibitem[\protect\citeauthoryear{{Donati}, {Semel}, {Carter}, {Rees} \&
  {Collier Cameron}}{{Donati} et~al.}{1997}]{donati97survey}
{Donati} J.-F.,  {Semel} M.,  {Carter} B.~D.,  {Rees} D.~E.,    {Collier
  Cameron} A.,  1997, MNRAS, 291, 658

\bibitem[\protect\citeauthoryear{{Donati}}{{Donati}}{2008}]{donati08}
{Donati} J.-F. e.~a.,  2008, MNRAS, In Press

\bibitem[\protect\citeauthoryear{Dunstone}{Dunstone}{2008}]{dunstone08thesis}
Dunstone N.,  2008, PhD thesis, University of St Andrews, St Andrews, Scotland
  (in preparation)

\bibitem[\protect\citeauthoryear{{Dunstone}, {Hussain}, {Cameron}, {Marsden},
  {Jardine}, {Stempels}, {Ramirez} \& {Donati}}{{Dunstone}
  et~al.}{2008}]{dunstone08}
{Dunstone} N.~J.,  {Hussain} G.~A.~J.,  {Cameron} A.~C.,  {Marsden} S.~C.,
  {Jardine} M.,  {Stempels} H.~C.,  {Ramirez} J.,    {Donati} J.-F.,  2008,
  MNRAS

\bibitem[\protect\citeauthoryear{{Hendry} \& {Mochnacki}}{{Hendry} \&
  {Mochnacki}}{2000}]{hendry00}
{Hendry} P.~D.,  {Mochnacki} S.~W.,  2000, ApJ, 531, 467

\bibitem[\protect\citeauthoryear{{Hussain}, {Allende Prieto}, {Saar} \&
  {Still}}{{Hussain} et~al.}{2006}]{hussain06}
{Hussain} G.~A.~J.,  {Allende Prieto} C.,  {Saar} S.~H.,    {Still} M.,  2006,
  MNRAS, 367, 1699

\bibitem[\protect\citeauthoryear{{Hussain}, {Donati}, {Collier Cameron} \&
  {Barnes}}{{Hussain} et~al.}{2000}]{hussain00}
{Hussain} G.~A.~J.,  {Donati} J.-F.,  {Collier Cameron} A.,    {Barnes} J.~R.,
  2000, MNRAS, 318, 961

\bibitem[\protect\citeauthoryear{{Lynden-Bell} \& {Boily}}{{Lynden-Bell} \&
  {Boily}}{1994}]{lyndenbell94}
{Lynden-Bell} D.,  {Boily} C.,  1994, MNRAS, 267, 146

\bibitem[\protect\citeauthoryear{{Marsden}, {Berdyugina}, {Donati}, {Eaton} \&
  {Williamson}}{{Marsden} et~al.}{2007}]{marsden07}
{Marsden} S.~C.,  {Berdyugina} S.~V.,  {Donati} J.-F.,  {Eaton} J.~A.,
  {Williamson} M.~H.,  2007, Astron.~Nachr., 328, 1047

\bibitem[\protect\citeauthoryear{{Montmerle}, {Grosso}, {Tsuboi} \&
  {Koyama}}{{Montmerle} et~al.}{2000}]{montmerle00}
{Montmerle} T.,  {Grosso} N.,  {Tsuboi} Y.,    {Koyama} K.,  2000, ApJ, 532,
  1097

\bibitem[\protect\citeauthoryear{{Morin}, {Donati}, {Forveille}, {Delfosse},
  {Dobler}, {Petit}, {Jardine}, {Cameron}, {Albert}, {Manset}, {Dintrans},
  {Chabrier} \& {Valenti}}{{Morin} et~al.}{2008}]{morin08}
{Morin} J.,  {Donati} J.-F.,  {Forveille} T.,  {Delfosse} X.,  {Dobler} W.,
  {Petit} P.,  {Jardine} M.~M.,  {Cameron} A.~C.,  {Albert} L.,  {Manset} N.,
  {Dintrans} B.,  {Chabrier} G.,    {Valenti} J.~A.,  2008, MNRAS, pp 26--+

\bibitem[\protect\citeauthoryear{{Petit}, {Donati}, {Oliveira}, {Auri{\`e}re},
  {Bagnulo}, {Landstreet}, {Ligni{\`e}res}, {L{\"u}ftinger}, {Marsden},
  {Mouillet}, {Paletou}, {Strasser}, {Toqu{\'e}} \& {Wade}}{{Petit}
  et~al.}{2004}]{petit04}
{Petit} P.,  {Donati} J.-F.,  {Oliveira} J.~M.,  {Auri{\`e}re} M.,  {Bagnulo}
  S.,  {Landstreet} J.~D.,  {Ligni{\`e}res} F.,  {L{\"u}ftinger} T.,  {Marsden}
  S.,  {Mouillet} D.,  {Paletou} F.,  {Strasser} S.,  {Toqu{\'e}} N.,    {Wade}
  G.~A.,  2004, MNRAS, 351, 826

\bibitem[\protect\citeauthoryear{{Petit}, {Donati}, {Wade}, {Landstreet},
  {Bagnulo}, {L{\"u}ftinger}, {Sigut}, {Shorlin}, {Strasser}, {Auri{\`e}re} \&
  {Oliveira}}{{Petit} et~al.}{2004}]{petit04hr1099}
{Petit} P.,  {Donati} J.-F.,  {Wade} G.~A.,  {Landstreet} J.~D.,  {Bagnulo} S.,
   {L{\"u}ftinger} T.,  {Sigut} T.~A.~A.,  {Shorlin} S.~L.~S.,  {Strasser} S.,
  {Auri{\`e}re} M.,    {Oliveira} J.~M.,  2004, MNRAS, 348, 1175

\bibitem[\protect\citeauthoryear{{Reiners}}{{Reiners}}{2006}]{reiners06}
{Reiners} A.,  2006, A\&A, 446, 267

\bibitem[\protect\citeauthoryear{{Scharlemann}}{{Scharlemann}}{1982}]{scharlem%
ann82}
{Scharlemann} E.~T.,  1982, ApJ, 253, 298

\bibitem[\protect\citeauthoryear{{Semel}, {Donati} \& {Rees}}{{Semel}
  et~al.}{1993}]{semel93}
{Semel} M.,  {Donati} J.-F.,    {Rees} D.~E.,  1993, A\&A, 278, 231

\bibitem[\protect\citeauthoryear{{Siess}, {Dufour} \& {Forestini}}{{Siess}
  et~al.}{2000}]{siess00}
{Siess} L.,  {Dufour} E.,    {Forestini} M.,  2000, A\&A, 358, 593

\bibitem[\protect\citeauthoryear{Simkin}{Simkin}{1974}]{simkin74}
Simkin S.,  1974, A\&A, 31, 129

\bibitem[\protect\citeauthoryear{{Strassmeier} \& {Rice}}{{Strassmeier} \&
  {Rice}}{2000}]{strass00}
{Strassmeier} K.~G.,  {Rice} J.~B.,  2000, A\&A, 360, 1019

\bibitem[\protect\citeauthoryear{{Strassmeier} \& {Rice}}{{Strassmeier} \&
  {Rice}}{2003}]{strassmeier03}
{Strassmeier} K.~G.,  {Rice} J.~B.,  2003, A\&A, 399, 315

\bibitem[\protect\citeauthoryear{Tonry \& Davis}{Tonry \&
  Davis}{1979}]{tonry79}
Tonry J.,  Davis M.,  1979, AJ, 84, 1511

\bibitem[\protect\citeauthoryear{{Vincent}, {Piskunov} \& {Tuominen}}{{Vincent}
  et~al.}{1993}]{vincent93}
{Vincent} A.,  {Piskunov} N.~E.,    {Tuominen} I.,  1993, A\&A, 278, 523

\bibitem[\protect\citeauthoryear{{Zahn}}{{Zahn}}{1977}]{zahn77}
{Zahn} J.-P.,  1977, A\&A, 57, 383

\end{thebibliography}

\end{document}